\documentclass[conference]{IEEEtran}

\usepackage{cite}
\usepackage{amsmath,amssymb,amsfonts}
\usepackage{graphicx}
\usepackage{textcomp}
\usepackage{xcolor}
\usepackage{booktabs}
\usepackage{multirow}
\usepackage{url}
\usepackage{xurl}  % better URL line breaking in two-column format
\usepackage{siunitx}

\makeatletter
\def\footnoterule{\relax%
  \kern-5pt % Pull line slightly up
  \hbox to \columnwidth{\vrule width 0.5\columnwidth height 0.4pt\hfill} % Line spans 50% of the column width
  \kern4.6pt % Space between line and footnote text
}
\makeatother

\begin{document}

\title{The Environmental Cost of Digital Sovereignty: Water, Energy, and Emissions Impacts of Sovereign AI Infrastructure in the Global South}

\author{
\IEEEauthorblockN{Muntaser Syed\IEEEauthorrefmark{1}, Marius C.\ Silaghi\IEEEauthorrefmark{1}, Sheikh Abujar\IEEEauthorrefmark{1}, Sharun Akter Khushbu\IEEEauthorrefmark{2}, Amal El Ahmad\IEEEauthorrefmark{3}}
\IEEEauthorblockA{\IEEEauthorrefmark{1}Florida Institute of Technology, Melbourne, FL, USA\\
msyed2011@my.fit.edu, msilaghi@fit.edu, sabujar2021@my.fit.edu}
\IEEEauthorblockA{\IEEEauthorrefmark{2}Daffodil International University, Dhaka, Bangladesh\\
sharun.cse@diu.edu.bd}
\IEEEauthorblockA{\IEEEauthorrefmark{3}UAE University, Al Ain, United Arab Emirates\\
700045933@uaeu.ac.ae}
}

\maketitle

\begin{abstract}
Sovereign AI has become a strategic priority across the Global South, with over \$200 billion in state-led commitments announced between 2024 and 2026. Yet the physical infrastructure that compute sovereignty demands, above all data centers, imposes water, energy, and carbon costs that fall hardest on countries least equipped to absorb them. This paper presents a comparative environmental stress analysis across four cases: the United Arab Emirates, Bangladesh, India, and Africa (with a focus on Kenya). Using publicly available water stress data, grid carbon intensity factors, and GPU power specifications, we model the water consumption, energy demand, and carbon emissions of hypothetical sovereign AI deployments under multiple cooling technology scenarios. We find that a 1,024-GPU cluster using evaporative cooling in the UAE would consume over 30 million liters of water annually in a country classified as ``extremely high'' water stress. In Bangladesh, sovereign AI policy documents call for centralized GPU procurement but do not address where to site data centers in a country where more than a fifth of the land floods in an average year and the power grid struggles to deliver reliable supply. We identify a sovereignty-sustainability trilemma in which no country can simultaneously maximize AI sovereignty, minimize environmental impact, and maintain affordable resource access for citizens. We propose design principles for environmentally responsible sovereign AI, including mandatory water usage effectiveness reporting, climate-vulnerability siting assessments, and a preference for frugal small language models over frontier pre-training in resource-constrained settings.\let\thefootnote\relax\footnotetext{© 2026 IEEE. Personal use of this material is permitted. Permission from IEEE must be obtained for all other uses, in any current or future media, including reprinting/republishing this material for advertising or promotional purposes, creating new collective works, for resale or redistribution to servers or lists, or reuse of any copyrighted component of this work in other works.}
\end{abstract}


%--------------------------------------------------------------
\section{Introduction}
\label{sec:introduction}

Between 2024 and 2026, governments across the Global South announced over \$200 billion in sovereign AI commitments. The United Arab Emirates launched Stargate UAE, a 5~GW data center campus backed by a consortium including OpenAI, Oracle, and NVIDIA \cite{stargateuae2025}. Saudi Arabia committed to 18,000 Blackwell GPUs through its HUMAIN initiative \cite{humain2025launch}. India deployed 34,381 GPUs across 14 cloud service providers under its IndiaAI Mission \cite{indiaai2024mission}. Forty-nine African nations endorsed the Kigali Declaration and a \$60 billion Africa AI Fund \cite{kigali2025declaration}. The strategic logic is clear: nations that depend on foreign providers for compute, data storage, and model access risk a new form of technological dependency.

The physical costs of this ambition are large and growing. The International Energy Agency projects that global data center electricity consumption will rise from 415~TWh in 2024 to 945 to 1,050~TWh by 2030 \cite{iea2025energy}. Water consumption by data centers in the Middle East and Africa alone is expected to increase from 119 billion liters to 426 billion liters over five years \cite{mordor2026mea}. A single 1~MW data center can consume 25.5 million liters of water annually, enough for the daily needs of 300,000 people \cite{restofworld2026water}. Gartner estimates that power shortages will restrict 40\% of AI data centers by 2027 \cite{gartner2025power}.

These burdens do not fall evenly. The countries now pursuing sovereign AI are among the world's most water-stressed, energy-constrained, and climate-vulnerable. Of 52 developing countries with active sovereign AI programs, 70.6\% face high overall water risk according to the World Resources Institute Aqueduct index \cite{wri2023aqueduct}. Sixteen countries sit in ``double jeopardy,'' combining high water stress (score $\geq 3.0$) with high grid carbon intensity ($\geq 400$~gCO$_2$/kWh), including all Gulf states, India, South Africa, and Egypt. Developing countries pursuing sovereign AI average grid carbon intensities 44\% higher than high-income countries (457 vs.\ 318~gCO$_2$/kWh) \cite{ember2025electricity}.

Yet most scholarship on sovereign AI focuses on geopolitics, economics, and cultural alignment \cite{singh2025sovereign, stanfordhai2026dilemma, brookings2025sovereignty}. Environmental impact is treated as an engineering footnote rather than a first-order policy variable. National AI strategy documents, including those of Bangladesh, the UAE, and most African Union member states, contain detailed provisions for data governance and compute procurement but little or no analysis of water sourcing, grid capacity, or emissions accounting \cite{bangladesh2026aipolicy, au2024aistrategy}.

This paper asks: what are the environmental trade-offs in water consumption, energy demand, carbon emissions, and resource competition that arise from deploying sovereign AI infrastructure in climate-vulnerable and resource-constrained developing countries, and how should these costs inform the design and siting of humanitarian AI systems?

We present a comparative environmental stress analysis across four cases selected to span the space of fiscal capacity and environmental vulnerability: the UAE, Bangladesh, India, and Africa (with Kenya as the primary focus). Using publicly available water stress data from WRI Aqueduct \cite{wri2023aqueduct}, grid carbon intensity factors from Ember \cite{ember2025electricity}, and GPU power specifications, we model the water consumption, energy demand, and carbon emissions of sovereign AI deployments under multiple cooling technology and PUE scenarios. We identify a sovereignty-sustainability trilemma: no country can simultaneously maximize AI sovereignty, minimize environmental impact, and maintain affordable resource access for citizens. We propose seven design principles for environmentally responsible sovereign AI and argue that frugal approaches (small language models, edge inference, parameter-efficient fine-tuning) represent the most defensible path for resource-constrained settings.

This paper responds to the GHTC 2026 theme of ``Technologies in Context'' by treating environmental context as inseparable from technology deployment decisions. A sovereign AI system that displaces agricultural water, destabilizes a fragile power grid, or increases carbon emissions in a climate-vulnerable country is not deployed in context, regardless of its technical performance.

%--------------------------------------------------------------
\section{Background and Definitions}
\label{sec:background}

\subsection{Defining Sovereign AI}

Sovereign AI is the capability to govern strategically important AI systems in line with national law and public interest, while retaining sufficient operational autonomy over data, compute, model adaptation, and safety controls \cite{singh2025sovereign}. Singh and Sengupta propose a four-pillar framework: Data (collection, storage, and governance under domestic jurisdiction), Compute (physical infrastructure for training and inference), Models (development, adaptation, and deployment of AI systems), and Norms (regulatory standards, safety protocols, and ethical guidelines) \cite{singh2025sovereign}. Sovereignty along these pillars is not binary. It ranges from full domestic control to managed interdependence with foreign providers \cite{brookings2025sovereignty}. The Stanford Human-Centered AI Institute distinguishes sovereign AI from adjacent concepts: data sovereignty concerns jurisdiction over personal and government data, cloud sovereignty concerns hosting location and access control, and digital sovereignty describes broader national control over information systems \cite{stanfordhai2026dilemma}. This paper focuses on compute sovereignty, because it is the pillar with the heaviest physical footprint.

\subsection{The Physical Reality of Compute Sovereignty}

Compute sovereignty requires domestic data centers. Data centers require power, cooling, water, and land. Unlike model development (which can build on open-weight foundations) or data governance (which is primarily legal and regulatory), compute sovereignty has an irreducible physical cost that scales with ambition.

An NVIDIA DGX H100 houses 8 GPUs and draws up to 10.2~kW at peak. A 256-GPU cluster (32 DGX nodes) requires roughly 0.33~MW of IT power. Applying a power usage effectiveness (PUE) ratio of 1.20 for a new hyperscale build yields 0.39~MW of total facility power. At PUE 1.80 (typical for enterprise-grade facilities in hot climates), total facility power rises to 0.59~MW. A 1,024-GPU cluster requires 1.31~MW of IT power and between 1.57~MW (PUE 1.20) and 2.35~MW (PUE 1.80) of facility power \cite{uptime2024survey}. The IEA reports a global average PUE of 1.41, with hyperscale operators achieving 1.14 and enterprise facilities averaging 1.92 \cite{iea2025energy}.

Each megawatt of facility power demands cooling. In hot climates, evaporative cooling systems consume 1.8 to 3.0 liters of water per kWh of IT load \cite{patterns2025carbon}. In arid regions, that water must come from somewhere: groundwater, desalination, or municipal supply. The power itself must come from the national grid, which in many developing countries runs on fossil fuels and already struggles to meet residential and industrial demand.

\subsection{The Sovereignty-Sustainability Tension}

This physical reality creates a tension that existing policy frameworks do not resolve. Nations face what we term a sovereignty-sustainability trilemma: the simultaneous pursuit of three objectives that prove mutually constraining.

The first objective is AI sovereignty itself, meaning domestic control over compute, data, and models. The second is environmental sustainability, meaning the minimization of water consumption, energy demand, and carbon emissions. The third is affordable resource access for citizens, meaning that water, electricity, and land remain available for agriculture, housing, and basic services rather than being redirected to data center operations.

No country in our analysis maximizes all three. The UAE maximizes sovereignty through massive capital deployment but at extreme environmental cost. Bangladesh aspires to sovereignty but lacks the infrastructure to pursue it without severe resource trade-offs. India achieves partial sovereignty with a growing environmental burden. Kenya achieves modest sovereignty with minimal environmental cost through renewable-anchored siting. The Tony Blair Institute and Brookings Institution have identified versions of this tension \cite{tbi2025sovereignty, brookings2025sovereignty}, but neither frames it in environmental terms. This paper provides that framing through a physics-grounded analysis of water, energy, and emissions at the facility level, building on the energy measurement foundations established by Syed et al.\ in the TOML framework \cite{syed2026toml, horowitz2014energy}.

%--------------------------------------------------------------
\section{Methodology}
\label{sec:methodology}

This paper employs a comparative multi-case environmental stress analysis. We selected four cases to represent distinct positions along two axes: fiscal capacity for sovereign AI investment (high vs.\ low) and environmental vulnerability (high vs.\ low). Table~\ref{tab:case_selection} shows the resulting matrix.

\begin{table}[htbp]
\caption{Case Selection Matrix}
\label{tab:case_selection}
\centering
\begin{tabular}{p{2.2cm}p{2.5cm}p{2.5cm}}
\toprule
 & \textbf{High env.\ vulnerability} & \textbf{Lower env.\ vulnerability} \\
\midrule
\textbf{High fiscal capacity} & UAE & India \\
\textbf{Low fiscal capacity} & Bangladesh & Africa (Kenya) \\
\bottomrule
\end{tabular}
\end{table}

We treat the Global South as a political and economic category, not a geographic one, following its common use for countries outside the established centers of technological power, which can include high-income states. The UAE qualifies: it builds sovereign AI from outside the incumbent compute powers and under extreme water stress, placing it at the center of our trilemma. The classification is contested, and we include the UAE to span the range of fiscal capacity under a shared environmental constraint.

The UAE represents a wealthy petrostate deploying frontier-scale compute in an extreme desert climate. Bangladesh represents a low-income country with ambitious AI policy but severe climate exposure and infrastructure deficits. India represents large-scale, state-backed AI investment on a coal-heavy grid with significant regional water stress. Africa (with Kenya as the primary focus) represents a continental approach anchored in renewable energy and regional resource pooling.

For each case, we assess sovereign AI deployment plans against five environmental stress vectors: (1) water consumption and scarcity, (2) energy demand and grid capacity, (3) carbon emissions, (4) land use and siting constraints, and (5) resource competition with human needs, including agricultural irrigation, residential water supply, and household electrification.

The quantitative analysis models hypothetical GPU clusters at two scales (256 and 1,024 GPUs) using published NVIDIA DGX H100 power specifications (10.2~kW per 8-GPU node). We estimate annual water consumption under three cooling regimes (evaporative, hybrid, and immersion) using water usage effectiveness (WUE) values from peer-reviewed literature \cite{patterns2025carbon}, adjusted for local climate conditions. We estimate annual carbon emissions by multiplying facility-level electricity consumption by country-specific grid carbon intensity factors from Ember \cite{ember2025electricity}. We model two PUE scenarios: 1.20 (new hyperscale construction) and 1.80 (enterprise-grade facilities in hot climates), based on Uptime Institute benchmarks \cite{uptime2024survey}. Because ML workloads are predominantly memory-bound, with memory operations consuming 60 to 90\% of total energy \cite{syed2026toml, horowitz2014energy}, GPU TDP-based power estimates represent conservative lower bounds on actual energy consumption.

Water stress scores are drawn from the WRI Aqueduct 4.0 Water Risk Atlas, which covers 228 countries across multiple risk dimensions including baseline water stress, riverine flood risk, coastal flood risk, and drought risk \cite{wri2023aqueduct}. Climate vulnerability data draws on World Bank risk assessments \cite{worldbank2021bangladesh} and IPCC classifications. National AI policy analysis is based on primary government documents \cite{bangladesh2026aipolicy, indiaai2024mission, au2024aistrategy} and UNESCO readiness assessments \cite{unesco2025bangladeshram}.

Several limitations apply. Capital expenditure data for sovereign AI projects is often incomplete or promotional. Water consumption and carbon emissions are site-specific, and our models use national or regional averages rather than facility-level measurements. The analysis produces order-of-magnitude estimates suitable for cross-country comparison, not engineering-grade predictions for individual deployments.

Our estimates are robust to the three driving parameters. PUE across the modeled 1.20 to 1.80 range scales energy and carbon linearly, so the Bangladesh cluster spans 9,554 to 14,330 tons of CO$_2$ per year. Evaporative WUE of 1.8 to 3.0 liters per kWh \cite{patterns2025carbon} places the UAE water estimate between 20.6 and 34.3 million liters, and we report the upper bound for its arid climate. Grid carbon intensity dominates: the 95.4 to 696.1~gCO$_2$/kWh spread is a factor of 7.3, exceeding the combined PUE and WUE range, so the country ranking holds across the full parameter space.

%--------------------------------------------------------------
\section{Environmental Stress Analysis}
\label{sec:analysis}

\subsection{Water Consumption and Scarcity}
\label{subsec:water}

\begin{figure}[t]
\centering
\includegraphics[width=\columnwidth]{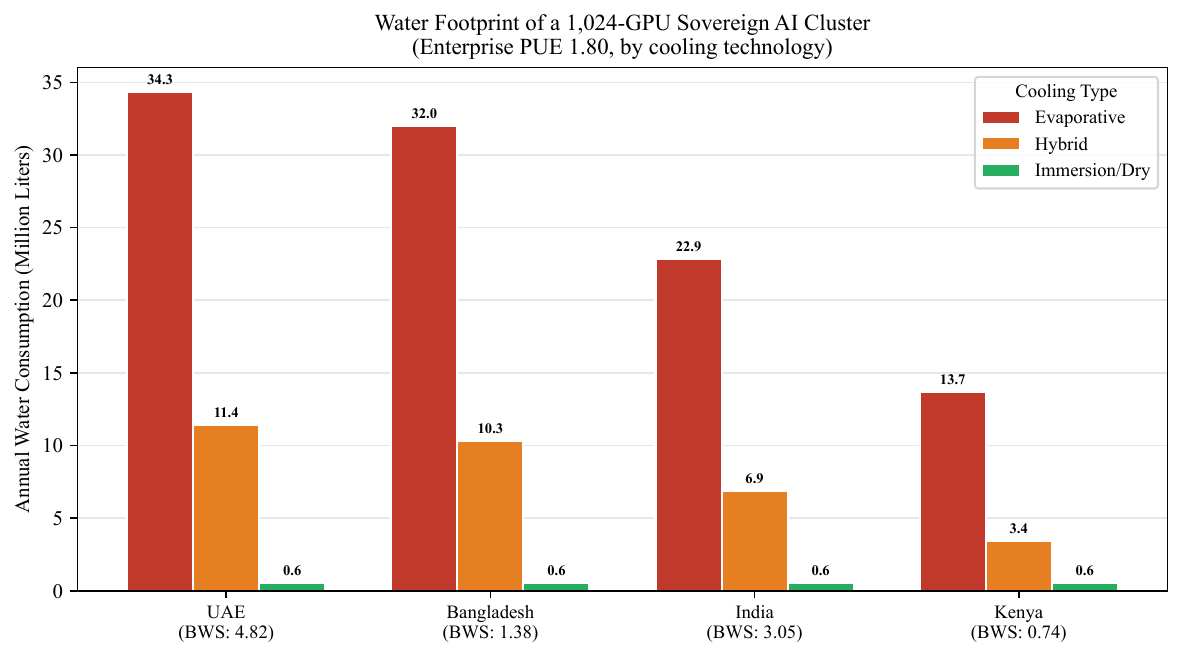}
\caption{Annual water consumption for a 1,024-GPU cluster by country and cooling technology (PUE 1.80). Evaporative cooling in the UAE consumes 34.3 million liters per year at a water stress score of 4.82/5.0.}
\label{fig:water}
\end{figure}

Data center cooling drives water consumption, with evaporative systems consuming 1.8 to 3.0 liters per kWh of IT load depending on climate and design \cite{patterns2025carbon}. Our model estimates annual water consumption for a 1,024-GPU cluster (1.31~MW IT load) under three cooling regimes. Results vary by more than two orders of magnitude depending on technology choice and location. In the UAE, evaporative cooling consumes 34.3 million liters per year, equivalent to the daily water needs of 940 people (Fig.~\ref{fig:water}). Bangladesh consumes 32.0 million liters under the same configuration. India consumes 22.9 million liters; Kenya consumes 13.7 million liters. Switching to immersion cooling reduces UAE water consumption to 0.57 million liters, a 98.3\% reduction.

These numbers become alarming when placed against national water stress. The WRI Aqueduct index assigns the UAE a baseline water stress score of 4.82 out of 5.0, classified as ``Extremely High'' \cite{wri2023aqueduct}. India scores 3.05 (``High''), with 43.4\% of sub-regions in the ``Extremely High'' category. Bangladesh presents a paradox: its average water stress score is a moderate 1.38, but it simultaneously scores High or above on riverine flood risk (4.04), coastal flood risk (3.19), and drought risk (3.54). It is the only developing country with active sovereign AI programs that exceeds 3.0 on three distinct water risk dimensions. The question for Bangladesh is not whether water exists, but whether any of it is safe, stable, and accessible at data center sites. Arsenic contamination affects 35 to 77 million people \cite{smith2000arsenic}, and in an average year more than a fifth of the country's land floods, reaching up to 70 percent in extreme flood years \cite{worldbank2021bangladesh}.

\subsection{Energy Demand and Grid Capacity}
\label{subsec:energy}

AI training workloads consume 7 to 8 times more energy than standard compute tasks of comparable duration. A single AI query can use 1,000 times more electricity than a conventional web search \cite{enkiai2026grid}. Africa, the Middle East, and Latin America combined receive less than 4\% of new global data center capacity \cite{iea2025energy}.

Our 1,024-GPU cluster requires 1.31~MW of IT power. At PUE 1.80, total facility demand rises to 2.35~MW. This is a modest cluster by hyperscale standards, yet it places significant load on grids that already struggle with reliability. In Bangladesh, the grid reaches nearly the entire population, but supply reliability remains weak \cite{worldbank2022electricity}. Mobile broadband averages 9.2~Mbps against a global average of 64.2~Mbps \cite{unesco2025bangladeshram}. The national AI policy calls for centralized GPU procurement across all government agencies but does not specify how the grid will absorb the additional load \cite{bangladesh2026aipolicy}.

In Africa, one in two people lacks reliable electricity. Each megawatt dedicated to a data center is a megawatt not available for household electrification, hospital power, or agricultural processing. The Stargate UAE campus alone plans for 5~GW of capacity \cite{stargateuae2025}, roughly equivalent to the entire installed generation capacity of Kenya.

\subsection{Carbon Emissions}
\label{subsec:carbon}

\begin{figure}[t]
\centering
\includegraphics[width=\columnwidth]{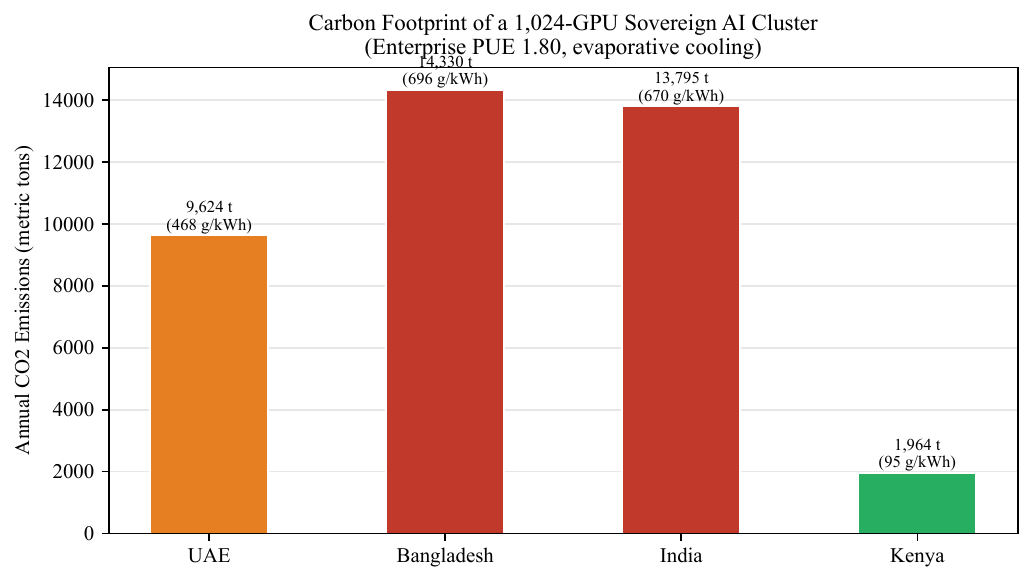}
\caption{Annual CO$_2$ emissions for a 1,024-GPU cluster by country (PUE 1.80). Grid carbon intensity ranges from 95.4~gCO$_2$/kWh (Kenya) to 696~gCO$_2$/kWh (Bangladesh).}
\label{fig:carbon_country}
\end{figure}

The carbon cost of sovereign AI depends almost entirely on grid mix. A 1,024-GPU cluster running at PUE 1.80 emits 14,330 tons of CO$_2$ per year on the Bangladesh grid (696~gCO$_2$/kWh), 13,795 tons on the Indian grid (670~gCO$_2$/kWh), and 9,624 tons on the UAE grid (467.5~gCO$_2$/kWh). The same cluster in Kenya emits 1,964 tons (95.4~gCO$_2$/kWh). Bangladesh emits 7.3 times more CO$_2$ than Kenya for identical computation (Fig.~\ref{fig:carbon_country}).

\begin{figure}[t]
\centering
\includegraphics[width=\columnwidth]{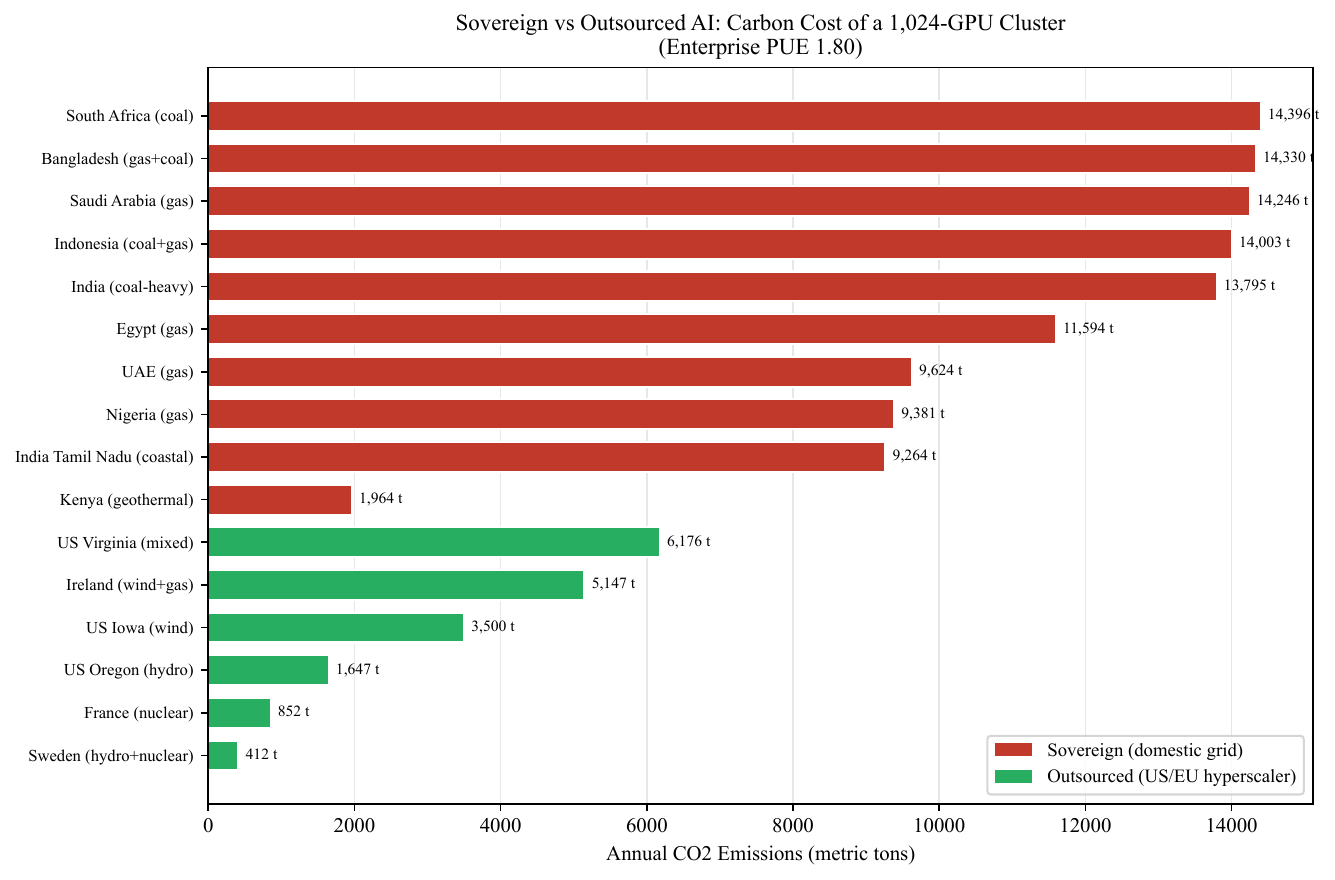}
\caption{Annual CO$_2$ emissions for a 1,024-GPU cluster: sovereign deployments (domestic grid) vs.\ outsourced deployments (US and EU hyperscaler regions). The same computation in Sweden produces 35 times less CO$_2$ than in Bangladesh.}
\label{fig:carbon}
\end{figure}

The comparison with outsourced alternatives is striking (Fig.~\ref{fig:carbon}). The same workload running on Swedish hydro and nuclear power (20~gCO$_2$/kWh) emits 412 tons per year. Running on Oregon hydropower (80~gCO$_2$/kWh) emits 1,647 tons. In pure carbon terms, outsourcing computation to a clean-grid hyperscaler produces a fraction of the emissions of a sovereign deployment on a coal- or gas-heavy domestic grid. This does not settle the sovereignty question, but it quantifies the environmental price of keeping computation at home.

Diesel backup generators compound the problem. In countries with unreliable grids, data centers maintain diesel generators for continuity. These generators can emit 2 to 3 times the CO$_2$ per kWh of even the dirtiest national grids, and their use is rarely reported in facility-level emissions data.

\subsection{Resource Competition}
\label{subsec:resource_competition}

These costs compete directly with human needs. Water used for cooling is unavailable for agriculture, drinking, or sanitation, and power for GPU clusters does not reach households, clinics, or schools.

In India, 271 data centers consumed roughly 150 billion liters of water in 2024, and that figure is expected to double by 2030. These facilities draw from the same basins that supply irrigation for a country where agriculture employs over 40\% of the workforce. In the UAE, per capita daily water consumption already exceeds 500 liters, nearly three times the global average. Desalination provides 42\% of drinking water, and the feedback loop is direct: more data centers require more cooling water, which requires more desalination, which requires more energy, which produces more emissions \cite{mecouncil2025desert}.

In Bangladesh, the competition is even starker. The country has 1,100 people per square kilometer, making it the most densely populated large country on Earth. Available land is contested between housing, agriculture, flood mitigation infrastructure, and now data centers. Power rationing is already common. The addition of GPU clusters to a fragile grid creates a zero-sum allocation problem that the national AI policy does not address \cite{bangladesh2026aipolicy}.

In sub-Saharan Africa, 600 million people lack electricity access. A \$300 million investment by the U.S.\ International Development Finance Corporation in Africa Data Centres created 31.3~MW of capacity and roughly 100 direct jobs. The same power could alternatively serve thousands of homes. The resource trade-off is not hypothetical; it is the daily reality of infrastructure planning in energy-poor countries.

%--------------------------------------------------------------
\section{Case Studies}
\label{sec:cases}

\subsection{UAE: Frontier Sovereignty Under Desert Stress}
\label{subsec:uae}

The Stargate UAE campus plans for up to 5~GW of data center capacity with 2.5 million NVIDIA chips \cite{stargateuae2025}. Saudi Arabia's HUMAIN initiative has procured 18,000 Blackwell GPUs \cite{humain2025launch}. The UAE grid runs on 68.3\% natural gas and 22.9\% nuclear, producing 467.5~gCO$_2$/kWh \cite{ember2025electricity}. Our model estimates that a 1,024-GPU cluster with evaporative cooling in the UAE would consume 34.3 million liters of water per year and emit 9,624 tons of CO$_2$ at PUE 1.80.

\begin{figure}[t]
\centering
\includegraphics[width=\columnwidth]{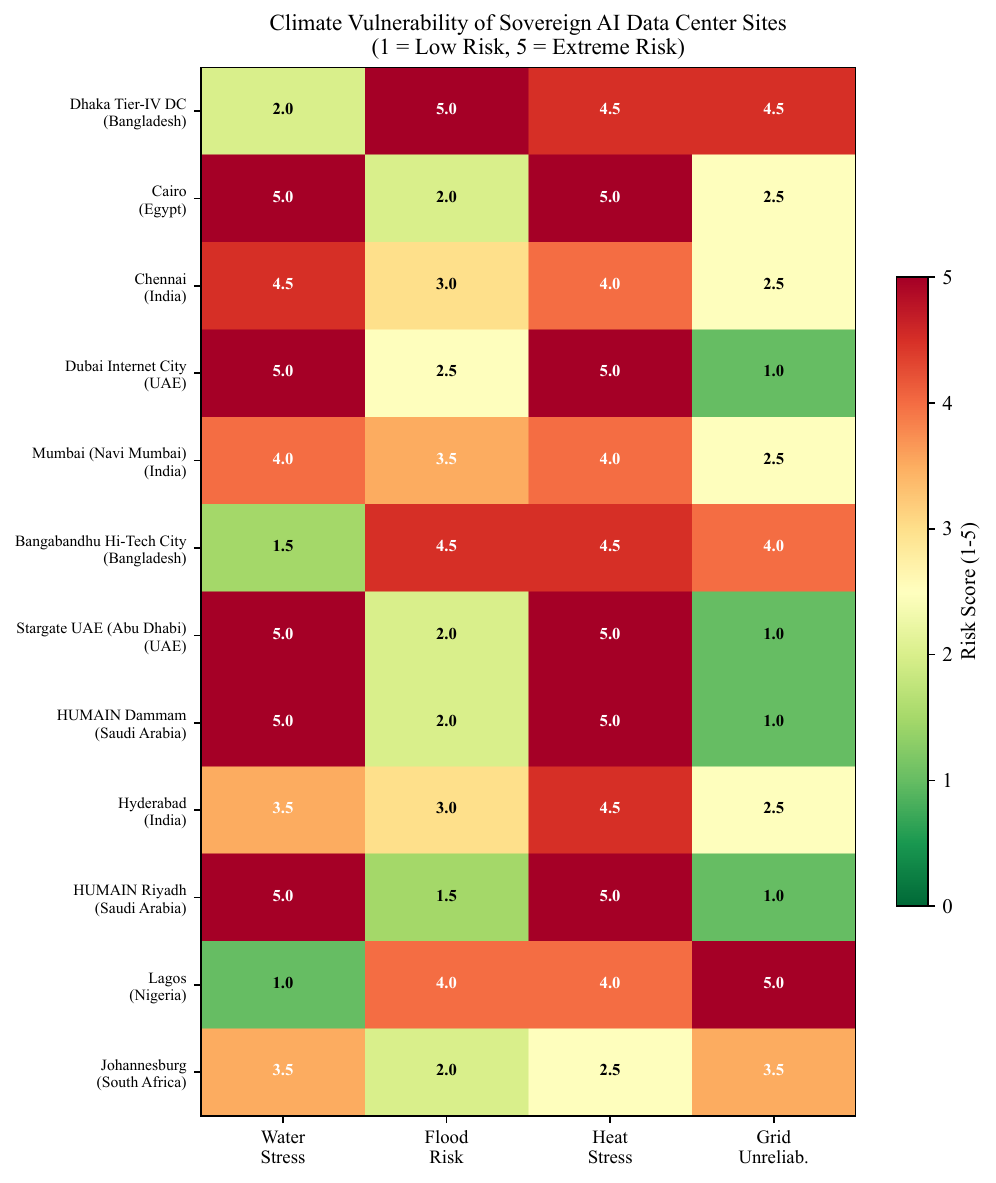}
\caption{Climate vulnerability scores for 15 actual and planned data center sites across five risk dimensions. Dhaka scores highest overall (3.88); Nairobi Konza scores lowest (1.93).}
\label{fig:heatmap}
\end{figure}

The water problem is structural. WRI Aqueduct assigns the UAE a baseline water stress score of 4.82 out of 5.0, with 96.4\% of sub-regions classified as high or extremely high stress \cite{wri2023aqueduct}. Our vulnerability assessment scores the Stargate Abu Dhabi site at 5.0 for water stress and 5.0 for heat stress, with peak temperatures reaching 50\textdegree C (Fig.~\ref{fig:heatmap}). Desalination provides 42\% of drinking water, and the region accounts for 90\% of global thermal energy used in desalination \cite{mecouncil2025desert}.

\begin{figure}[t]
\centering
\includegraphics[width=\columnwidth]{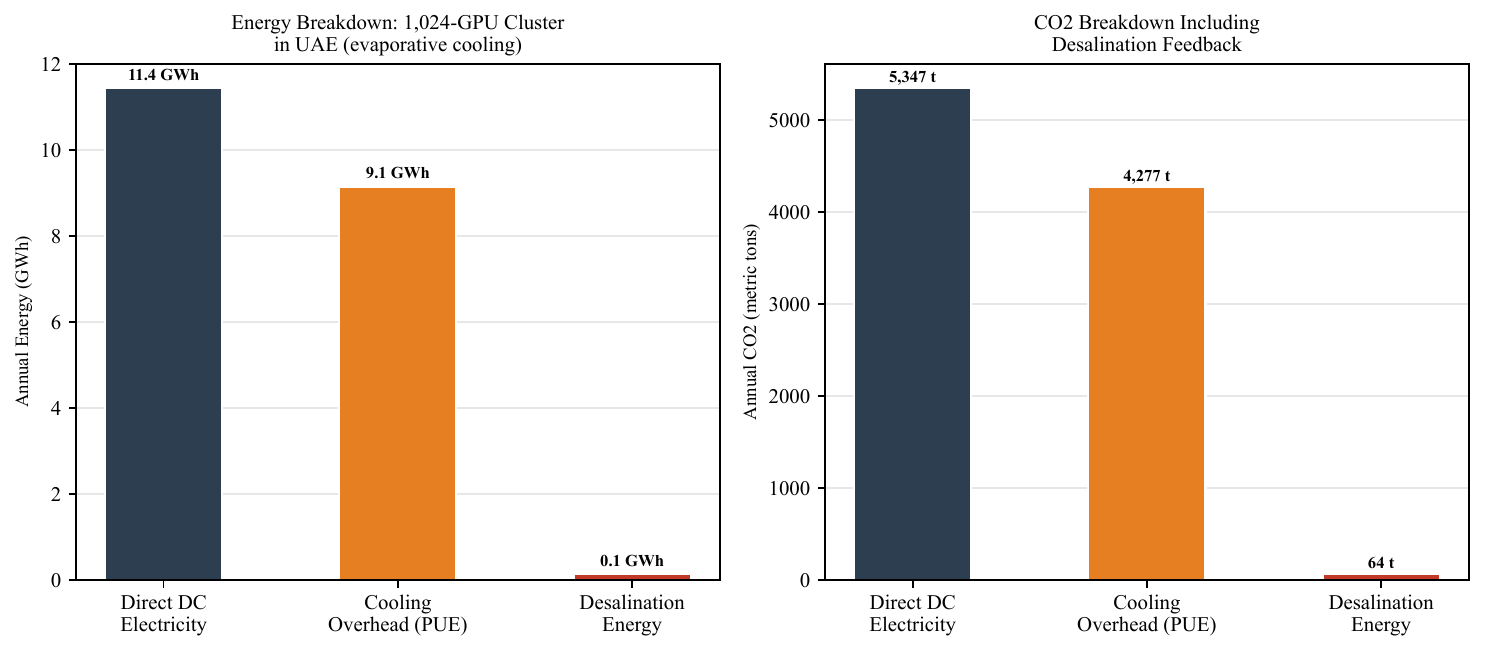}
\caption{UAE desalination feedback loop: energy and CO$_2$ breakdown for a 1,024-GPU sovereign AI deployment, showing the additional overhead from desalination-sourced cooling water.}
\label{fig:desal}
\end{figure}

This creates a feedback loop: more AI compute requires more cooling water, which requires more desalination, which requires more energy, which produces more emissions (Fig.~\ref{fig:desal}). Immersion cooling offers a partial solution, reducing water consumption by 98.3\% in our model. Companies like XDS Datacentres have deployed full immersion cooling in Riyadh and Jeddah with zero water for cooling. The UAE case shows that unlimited capital can buy compute sovereignty, but it cannot eliminate the physical constraints of a desert climate.

\subsection{Bangladesh: Sovereignty Without Infrastructure}
\label{subsec:bangladesh}

Bangladesh's National AI Policy 2026-2030 calls for centralized GPU procurement across all government agencies, mandatory data localization for health, biometric, financial, and security data, and sovereign cloud infrastructure \cite{bangladesh2026aipolicy}. The UNESCO AI Readiness Assessment identifies 15 priority gaps, including GPU scarcity, outdated curricula, and AI ethics capacity described as ``nearly absent'' \cite{unesco2025bangladeshram}. Digital literacy stands at 8\%. Internet access reaches 44.5\% of the population.

The environmental profile is the most complex in our study. Bangladesh is the only developing country with sovereign AI programs in our merged dataset that scores High or above on three simultaneous water risk dimensions: riverine flood risk (4.04), coastal flood risk (3.19), and drought risk (3.54) \cite{wri2023aqueduct}. About 60 percent of the population lives less than five meters above sea level \cite{worldbank2021bangladesh}. Arsenic contamination in groundwater affects 35 to 77 million people \cite{smith2000arsenic}. Our vulnerability overlay scores the Dhaka Tier-IV data center site at 3.88 composite risk, the highest of all 15 sites assessed, with a riverine flood score of 5.0 and grid unreliability at 4.5.

The grid compounds the problem. Bangladesh's electricity mix is 97.9\% fossil fuel (64.3\% gas, 21.5\% coal), producing 696.1~gCO$_2$/kWh, the highest carbon intensity among our four cases \cite{ember2025electricity}. A 1,024-GPU cluster at PUE 1.80 would emit 14,330 tons of CO$_2$ per year, 7.3 times more than the same cluster in Kenya. The policy document does not address where to physically site data centers, how to source clean water for cooling, or how the grid will absorb additional load. Bangladesh represents the case where sovereignty ambitions collide most directly with physical reality.

\subsection{India: Layered Sovereignty at Scale}
\label{subsec:india}

India's IndiaAI Mission has deployed 34,381 GPUs across 14 cloud service providers at a subsidized rate of Rs~65 per GPU-hour \cite{indiaai2024mission}. The India Semiconductor Mission has committed Rs~76,000 crore (roughly \$9.1 billion) in fiscal support for domestic fabrication. India pursues what can be described as layered sovereignty: strict control over data governance and standards, combined with managed dependence on imported hardware.

The environmental burden is proportional to the ambition. India's grid carbon intensity is 670.1~gCO$_2$/kWh, driven by a 70.8\% coal share \cite{ember2025electricity}. Average water stress scores 3.05 (``High''), with 43.4\% of sub-regions in the ``Extremely High'' category \cite{wri2023aqueduct}. Our vulnerability assessment scores Chennai at 3.6 composite risk (water stress 4.5) and Mumbai at 3.58 (water stress 4.0). A 1,024-GPU cluster at PUE 1.80 would consume 22.9 million liters of water and emit 13,795 tons of CO$_2$ per year.

India's scale, however, enables mitigation strategies that smaller countries cannot pursue. Coastal siting along the Tamil Nadu and Gujarat coasts positions data centers near offshore wind and solar generation, reducing effective grid carbon intensity to roughly 450~gCO$_2$/kWh in those regions. India has also invested in frugal AI development: models like Sarvam-30B and Sarvam-105B were trained on 4,096 H100 GPUs at roughly \$30 million, an order of magnitude below frontier costs. India demonstrates that sovereign AI at scale is environmentally costly but not inevitably so, provided siting and model choices are made with environmental criteria in view.

\subsection{Africa: Negotiated Interdependence and the Geothermal Exception}
\label{subsec:africa}

Africa's approach to AI sovereignty differs structurally from the national models above. The Kigali Declaration, endorsed by 49 nations in April 2025, established a \$60 billion Africa AI Fund and a continental governance framework \cite{kigali2025declaration}. The African Union's Continental AI Strategy emphasizes shared regional infrastructure rather than 55 separate national stacks \cite{au2024aistrategy}. Cassava Technologies and NVIDIA have partnered on a \$700 million initiative to build AI data centers across five countries \cite{cassava2025nvidia}. The continent holds less than 2\% of global data center capacity and roughly 15,000 GPUs, projected to grow modestly by 2028.

Kenya is the standout case. Its grid is 90\% clean energy, with 20.9\% hydropower, 16.4\% wind, and 4.6\% solar; the remainder is primarily geothermal \cite{ember2025electricity}. Grid carbon intensity is 95.4~gCO$_2$/kWh, the lowest of any country in our study. Our model estimates that a 1,024-GPU cluster in Kenya would emit 1,964 tons of CO$_2$ per year, compared to 14,330 tons for the same cluster in Bangladesh. The Nairobi Konza Technopolis data center site scores 1.93 on our composite vulnerability index, the lowest of all 15 sites assessed, with water stress at 1.5 and heat stress at 1.5.

A \$1 billion geothermal-powered data center backed by Microsoft and G42 is under development, and the UNDP timbuktoo initiative has launched AI Compute Nodes on renewable micro-grids in Kenya, Rwanda, and South Africa. Kenya demonstrates that siting decisions driven by renewable energy access can reduce the environmental cost of sovereign AI by an order of magnitude. The constraint is that this model produces modest compute capacity. Africa's total continental GPU pool is smaller than a single planned UAE campus. The trade-off between environmental responsibility and competitive compute power is starkest on this continent.

%--------------------------------------------------------------
\section{Discussion}
\label{sec:discussion}

\subsection{The Sovereignty-Sustainability Trilemma}

The four cases reveal a structural tension. No country simultaneously maximizes AI sovereignty, minimizes environmental impact, and maintains affordable resource access for citizens.

The UAE maximizes sovereignty through capital-intensive deployment but at extreme environmental cost: 34.3 million liters of water per year for a single 1,024-GPU cluster in a country with a water stress score of 4.82/5.0. Bangladesh aspires to sovereignty through policy (centralized GPU procurement, mandatory data localization) but lacks the infrastructure to execute without severe environmental consequences: the Dhaka data center site scores 3.88 on our composite vulnerability index, the highest of 15 sites assessed, and the grid produces 696.1~gCO$_2$/kWh. India achieves partial sovereignty at scale but with growing absolute environmental burden: 13,795 tons of CO$_2$ per year from a single cluster on a 70.8\% coal grid. Kenya achieves modest sovereignty with minimal environmental cost (1,964 tons CO$_2$, composite vulnerability 1.93) but produces only a fraction of the compute capacity available to the other cases.

This trilemma is not a temporary engineering problem awaiting a technical fix. It reflects the physical constraints of deploying energy- and water-intensive infrastructure in climate-vulnerable, resource-constrained settings. Policy frameworks that treat sovereignty as a binary (sovereign vs.\ dependent) miss the environmental costs embedded in that choice.

\subsection{Frugal AI as the Environmentally Defensible Path}

\begin{figure}[t]
\centering
\includegraphics[width=\columnwidth]{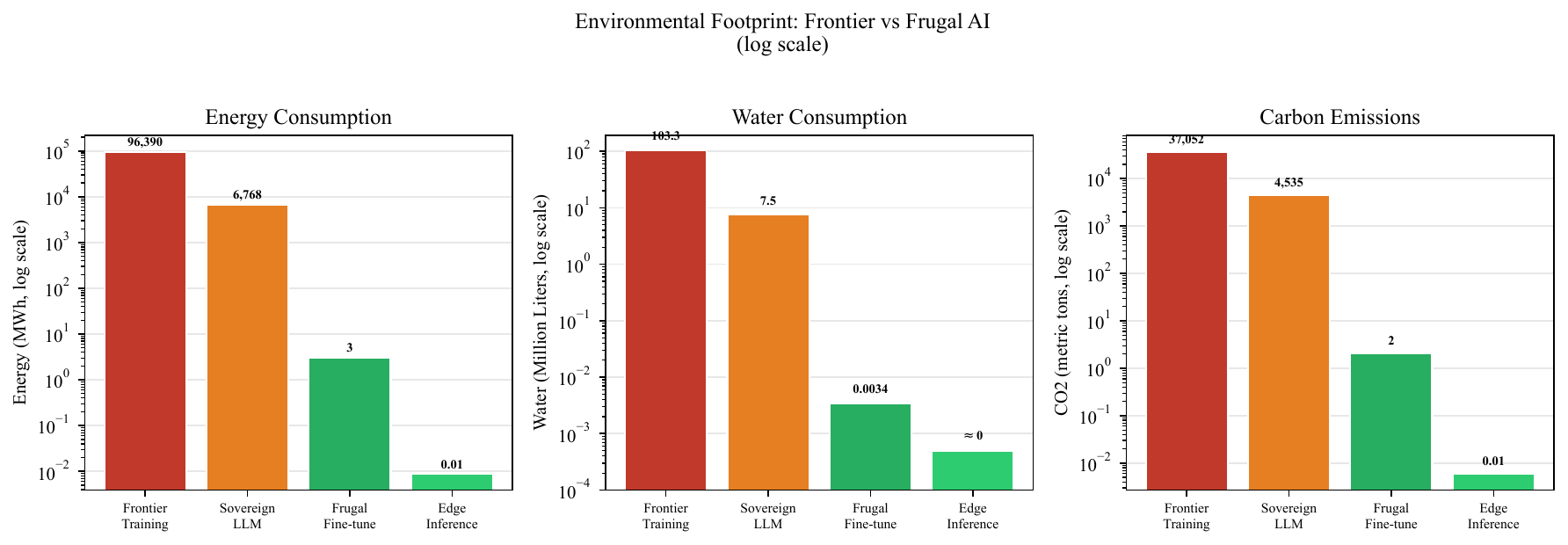}
\caption{Energy consumption across four AI deployment scenarios (log scale). Frontier pre-training consumes roughly 40,000 times more energy than frugal LoRA fine-tuning and nearly 8 million times more than edge inference. Edge inference consumes no data center cooling water, so its water bar sits at the axis floor and is marked as approximately zero.}
\label{fig:frugal}
\end{figure}

The environmental gap between frontier AI and frugal alternatives spans orders of magnitude. Table~\ref{tab:frugal} and Fig.~\ref{fig:frugal} summarize the comparison from our modeling.

\begin{table}[t]
\caption{Environmental Footprint by AI Development Scenario}
\label{tab:frugal}
\centering
\footnotesize
\begin{tabular}{lrrrr}
\toprule
\textbf{Scenario} & \textbf{GPUs} & \textbf{Energy} & \textbf{Water} & \textbf{CO$_2$} \\
 & & \textbf{(MWh)} & \textbf{(M liters)} & \textbf{(tons)} \\
\midrule
Frontier pre-training & 25,000 & 68,850 & 103.3 & 37,052 \\
Sovereign LLM & 4,096 & 3,760 & 7.5 & 4,535 \\
Frugal fine-tuning & 8 & 1.7 & 0.003 & 2.1 \\
Edge inference & 0 & 0.009 & 0 & 0.006 \\
\bottomrule
\end{tabular}
\end{table}

Frontier pre-training (25,000 GPUs for 90 days) consumes 68,850~MWh of IT energy, 103.3 million liters of water, and produces 37,052 tons of CO$_2$. Frugal fine-tuning using LoRA on 8 GPUs for 7 days consumes 1.7~MWh, 3,427 liters of water, and produces 2.1 tons of CO$_2$. The energy ratio is roughly 40,000 to 1. Edge inference on a smartphone (as demonstrated by Lelapa AI's InkubaLM at 400 million parameters) requires 8.8~kWh per year and produces 5.9~kg of CO$_2$, with zero data center water consumption.

Frugal AI reduces all five environmental stress vectors simultaneously: less water, less power, fewer emissions, less land, and less resource competition with human needs. For countries in the lower-left quadrant of our case selection matrix (low fiscal capacity, high environmental vulnerability), frugal approaches are not a compromise. They are the only environmentally defensible path to sovereign AI capability.

\subsection{Designing for Local Environmental Monitoring}

These results translate into concrete guidance for humanitarian and public health programs. The decisive choice is architectural. A cloud-dependent service sends every query to a remote data center, carrying that facility's water and energy cost and assuming broadband that often does not exist, as in Bangladesh at 9.2~Mbps \cite{unesco2025bangladeshram}. An offline design keeps computation local: small models on a clinic tablet, or SMS and USSD utilities on basic phones, need no connectivity, carry near-zero water and energy at the point of use, and keep working through the outages common in these settings.

National averages cannot characterize one small deployment, so local actors need a method using field-obtainable inputs. Applying our model at small scale, multiply device power in kilowatts by annual operating hours for energy in kWh, then by a local emission factor: grid intensity on grid, two to three times that for a diesel generator, or near zero for solar. Cooling water is negligible for edge workloads. In a temporary refugee settlement on diesel or solar, this shows a solar-powered SMS or edge service approaches zero marginal water and emissions, while a cloud service imports a distant data center's footprint plus the diesel cost of keeping equipment online.

\subsection{Design Principles for Environmentally Responsible Sovereign AI}

Based on the analysis above, we propose seven design principles:

\begin{enumerate}
\item \textbf{Mandatory WUE reporting.} All publicly funded sovereign compute should report Water Usage Effectiveness alongside PUE, enabling transparent comparison across facilities and cooling technologies.
\item \textbf{Climate-vulnerability siting assessments.} Data center construction should require assessment of flood risk, drought risk, sea level rise, and heat stress at the proposed site. Our vulnerability overlay (Fig.~\ref{fig:heatmap}) provides a template.
\item \textbf{Renewable-first compute.} Sovereign AI deployments should prioritize locations with existing or planned renewable generation. Kenya's geothermal model produces 7.3 times less CO$_2$ per unit of computation than Bangladesh's gas-and-coal grid.
\item \textbf{Frugal model preference.} Resource-constrained countries should default to small language model fine-tuning and edge inference rather than frontier pre-training. The energy difference is four orders of magnitude.
\item \textbf{Regional pooled infrastructure.} Countries below a minimum compute-demand threshold should pursue shared regional facilities rather than duplicating national stacks. The African Union's continental approach offers a template.
\item \textbf{Environmental integration in AI strategy.} National AI policy documents should include environmental impact assessments covering water sourcing, grid capacity, and emissions projections. The Bangladesh AI Policy 2026-2030, which contains zero mention of environmental siting or water, illustrates the current gap \cite{bangladesh2026aipolicy}.
\item \textbf{Desalination-loop accounting.} For arid-region deployments, total environmental cost must include the energy and emissions overhead of producing cooling water through desalination.
\end{enumerate}

\subsection{Technologies in Context}

The GHTC 2026 theme asks engineers to consider technology in context. The same 1,024-GPU cluster emits 14,330 tons of CO$_2$ on the Bangladesh grid or 412 tons outsourced to Sweden, and consumes 34.3 million liters of water in the UAE or 0.57 million with immersion cooling. These order-of-magnitude gaps depend entirely on where and how sovereign AI is deployed. Environmental context is not an add-on to engineering decisions. It is the decision.

%--------------------------------------------------------------
\section{Conclusion}
\label{sec:conclusion}

Sovereign AI without environmental accounting is incomplete sovereignty. This paper has shown that the physical infrastructure required for compute sovereignty imposes water, energy, emissions, and land costs that fall disproportionately on the countries least able to absorb them.

Across four cases the pattern is consistent. The UAE buys frontier compute at extreme water cost (4.82/5.0, 34.3 million liters per cluster), Bangladesh pursues sovereignty by policy onto the highest-vulnerability site (3.88) and dirtiest grid (696.1~gCO$_2$/kWh), India scales on a 70.8\% coal grid (13,795 tons per cluster), and Kenya's geothermal grid (95.4~gCO$_2$/kWh) cuts emissions 7.3 times.

We have identified a sovereignty-sustainability trilemma: no country in our study simultaneously maximizes AI sovereignty, minimizes environmental impact, and maintains affordable resource access for citizens. Frugal AI approaches (parameter-efficient fine-tuning, small language models, edge inference) reduce energy consumption by four orders of magnitude relative to frontier pre-training and offer the most environmentally defensible path for resource-constrained countries.

Three directions for future work follow from this analysis. First, facility-level environmental audits of sovereign AI deployments would replace the order-of-magnitude estimates used here with measured water, energy, and emissions data. Second, organizations responsible for national AI readiness assessments (UNESCO, OECD, the African Union) should integrate environmental stress metrics, including water risk, grid carbon intensity, and climate vulnerability, into their evaluation frameworks. Third, the humanitarian engineering community should develop standards for environmentally responsible AI deployment that treat water usage effectiveness, renewable energy sourcing, and climate-vulnerability siting as first-order design constraints rather than optional reporting metrics.

The next wave of technology infrastructure need not repeat the resource extraction patterns of previous industrial revolutions. But avoiding that outcome requires treating environmental cost not as an externality to be managed after deployment, but as a design variable to be optimized before the first GPU is powered on.

%--------------------------------------------------------------
\section*{Acknowledgments}
AI-based tools were used for formatting, spelling, and grammar checking during manuscript preparation. All research design, data collection, analysis, and intellectual contributions are solely the work of the authors.

\bibliographystyle{IEEEtran}
\bibliography{references}

\end{document}